\documentclass[manuscript]{aastex}

\usepackage{graphicx}

\begin{document}

\title{IRAM-30m large scale survey of $^{12}$CO(2-1) and $^{13}$CO(2-1) emission in the Orion molecular cloud}

\author{%
O. Bern\'e \altaffilmark{1,2,3}
N. Marcelino\altaffilmark{4}
J. Cernicharo \altaffilmark{1}
}

\email{olivier.berne@irap.omp.eu}

\altaffiltext{1}{Centro de Astrobiolog'a (CSIC/INTA), Ctra. de Torrej—n a Ajalvir, km 4, 28850, Torrej—n de Ardoz, Madrid, Spain}
\altaffiltext{2}{Universit\'e de Toulouse; UPS-OMP; IRAP;  Toulouse, France}
\altaffiltext{3}{CNRS; IRAP; 9 Av. colonel Roche, BP 44346, F-31028 Toulouse cedex 4, France}
\altaffiltext{4}{NRAO, 520 Edgemont Road, Charlottesville, VA 22902, USA}

\begin{abstract}

{ Using the IRAM 30m telescope we have surveyed a $1\times0.8^{\circ}$ part of the 
Orion molecular cloud in the $^{12}$CO and $^{13}$CO (2-1) lines with a maximal spatial resolution of $\sim$11'' and 
spectral resolution of $\sim$ 0.4 km~s$^{-1}$. 
The cloud appears filamentary, clumpy and with a complex kinematical structure. We derive 
an estimated mass of the cloud of 7700 M$_{\sun}$ (half of which is found 
in regions with visual extinctions $A_V$ below $\sim$10) and a dynamical age for the nebula of the order of 0.2 Myrs. 
The energy balance suggests that magnetic fields play an important role in supporting the cloud, at large and small scales. 
According to our analysis, the turbulent kinetic energy in the molecular gas due to outflows is comparable 
to turbulent kinetic energy resulting from the interaction of the cloud with the HII region. This latter feedback
appears negative, i.e. the triggering of star formation by the HII region is inefficient in Orion. The reduced data 
as well as additional products such as the column density map are made available 
online\footnote{ \url{http://userpages.irap.omp.eu/~oberne/Olivier_Berne/Data}}.}

\end{abstract}


\section{Introduction}

The structure and dynamical properties of the interstellar medium in galaxies is deeply
influenced by the feedback of massive stars. Their ionizing photons (extreme UV, $E>13.6$ eV) 
 generate HII regions, which expand rapidly within their parental molecular cloud. This expansion
compresses molecular clouds and can trigger their gravitational collapse towards the formation 
of a new generation of low mass stars \citep{elm98, der10}. The interaction of the HII region with the surrounding cloud
is also a source of hydrodynamical instabilities \citep{spi54, fri54,bmc10}, turbulence \citep{elm04} and chemical mixing \citep{roy95, ber12}
in the interstellar medium. In addition, outflows from young stellar objects can also inject kinetic energy
in their parent molecular cloud \citep{bac96}.

\begin{figure*}
\begin{center}
\includegraphics[width=18cm]{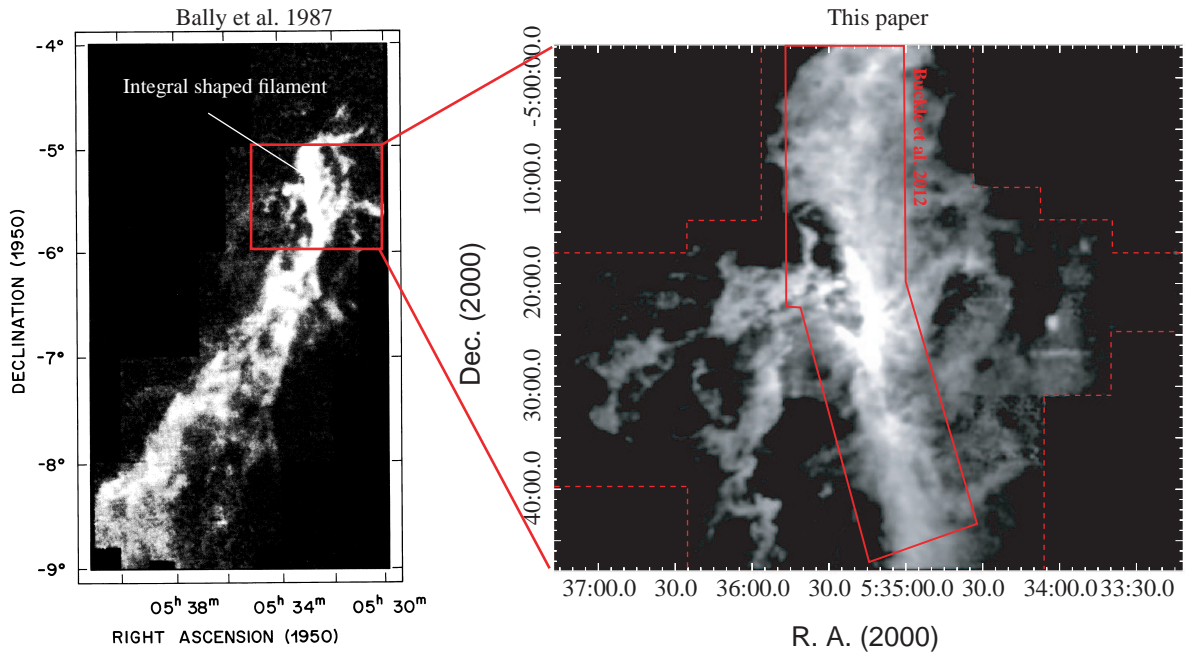}
\caption{ {  {\it Left:} Map of the $^{13}$CO (1-0) integrated intensity of the Orion A complex presented in \citet{bal87} with an angular resolution of 1'. 
{\it Right:} Map of the $^{13}$CO (2-1) integrated intensity (see Fig. 3 for details) obtained with the IRAM 30m telescope, with a maximal resolution of 11'', presented in this article. 
The limits of the mapped region are indicated with the red dashed line. The polygon with continuous red borders indicates the region mapped in
the CO (3-2) line by \citet{buc12}. Both maps are shown as gradations of gray scale intensities (see Fig. 3 for absolute scale).}
\label{fig_region}}
\end{center}
\end{figure*}

Given its proximity to us (414 pc, \citealt{men07}), the Orion nebula is one of the most studied regions of massive
star formation (see e.g. \citealt{gen89,bal08, mue08} for reviews). Traditionally, the ``Orion nebula" refers to the visible 
part of the region, the HII region, powered by the ionizing radiation of the 
Trapezium OB association. It is part of a much larger complex, refereed to as the Orion 
molecular cloud (OMC), itself lying at the border of the Eridanis super bubble.
Numerous studies have focused on specific objects in the Orion Nebula,
such as the famous Proplyds, the Orion Kleinmann-Low (KL) Nebula, or the Orion bar. Studies of the large scale 
structure of the nebula are however sparse, because of the difficulty 
to obtain observations covering large areas of the sky with sufficient angular resolution,
and because the interpretation of such datasets is challenging. \citet{bal87} reported
$^{13}$CO observation of the entire Orion molecular cloud, including the Nebula, with a resolution
of 1' (Fig.~\ref{fig_region}). This study was the first to reveal the filamentary structure of the cloud.
The Orion Nebula was found to lie at the center of an ``integral shaped filament" (ISF)
of molecular gas, itself part of a larger
filamentary structure extending from North to South over 4$^{\circ}$ (referred to as Orion A, 
see \citealt{bal87} and Fig.~\ref{fig_region}). Several studies have investigated the degree scale properties of 
the nebula in various molecular tracers. Using $^{12}$CO, $^{13}$CO and C$^{18}$O observations,
\citet{cas90} provided a column density and an H$_{2}$ density map of the ISF with an angular resolution 
of the order of 100''.
Using C$^{18}$O observations, \citet{dut91} identified clumps inside filaments distributed
periodically,  suggestive of externally triggered cloud collapse and fragmentation.
Recently, \citet{buc12} obtained a new survey in CO(3-2) along the ISF with a 15'' resolution (see their field of view in Fig.~\ref{fig_region}).
Their study concluded that the cloud is magnetically supported and wrapped in 
 a helical magnetic field. \cite{shi11} recently reported a survey of the Orion molecular cloud 
 in CO (1-0) with a spatial resolution of 20'' and a spectral resolution of 1 km~s$^{-1}$. 
 Using this dataset, the authors suggest that the formation of some clumps or protostars
 might have been triggered by the pressure exerted by the HII region \citep{shi11}.
  The ISF was observed by \citet{jon99} in the submillimeter
continuum. These authors found a strong correlation between the CO and cold dust continuum
emission. They also reported the presence of young stellar objects and pre-stellar clumps, mainly
found in the northern part of the ISF. Spitzer observation have shown further
evidence for this ongoing low mass star formation \citep{meg12}.
The HII region observed  at large scales with VLA at 330 MHz by \citet{sub01} shows a shell-like structure,
suggesting high pressures exerted on the Northern cloud, while the gas is escaping in the South
toward the general interstellar medium though cavities in the molecular gas. A similar idea was pointed out by \citet{gue08} who have
discovered the presence of an X-ray emitting plasma in thermodynamic equilibrium 
with the HII gas. Finally, the Veil of neutral gas situated in the foreground of the Orion nebula
was recently mapped at high angular and spectral resolution by \citet{vdw12}, and seemingly
confirms this idea that the HII region is in blister phase.

In this paper we present new observation of the $^{12}$CO and $^{13}$CO (2-1) rotational line at 1.3 mm
observed with the IRAM 30m telescope in Granada. This dataset has an unprecedented spatial
($\sim$11'') and spectral ($\sim$ 0.4 km~s$^{-1}$) resolution, together with a large field of view 
($1\times0.8^{\circ}$). This data is available at  \url{http://userpages.irap.omp.eu/~oberne/Olivier_Berne/Data}.

\section{Observations and data reduction}

{  The observations were conducted at the IRAM 30m telescope in Granada, Spain, in March, April
and October 2008. The data were obtained with the HERA receiver array, each polarization tuned to
$^{12}$CO and $^{13}$CO lines (230.5 and 220.4 GHz respectively) with a spatial beamwidth of 11'', 
 and main beam efficiencies of 0.524 and 0.545 for $^{12}$CO and $^{13}$CO respectively. 
 We used the versatile spectrometer VESPA  as a backend providing 320 kHz of spectral resolution, 
 (0.4 km~s$^{-1}$). We observed using the On-The-Fly (OTF) mapping mode, with 5'' of data sampling in 
 right ascension, and with steps of 12''  in declination. { Since the beam of the telescope is $\sim$11'',
 this implies that the resolution is equal to the telescope beam in right ascension, but is a factor of two 
 larger in declination.}  Intensity calibration was performed every 3-4min
 using two absorbers at different temperatures, resulting in system temperatures between 300 to 500 K. 
 Atmospheric opacities were obtained from the measurement of the sky emissivity and the use of the ATM code \citep{cer85,par01}, 
 and found to be between 0.1 and 0.2 corresponding to 1 to 3 mm of precipitable water vapor. 
 Focus and pointing were checked every 3 hours and 1.5--2 hours respectively on intense nearby sources.
 The map is centered at the position of the infrared source IRc2 ($\alpha$=05h35m14.5s, $\delta$=-05d22m29.3s $/$
 $\alpha$=83.81037$^{\circ}$, $\delta$=-05.3748$^{\circ}$ ).
The reference position which is free of CO emission was located at an offset (-3600'',-1800'').
Data reduction was performed using the GILDAS software\footnote{http://www.iram.fr/IRAMFR/GILDAS}. 
Several corrections were applied to the data:  one of the problems concerns the
combination of submaps obtained over different observing sessions with different
 intensity calibration, weather conditions and pointing errors. Position switching observations (PSW) 
 were performed at the position of IRc2 before each OTF submap and these were used afterwards to correct 
 for such differences between submaps. Relative errors in pointing corrections were obtained 
 comparing these PSW observations with the 4$''$ spacing $^{12}$CO  map observed within the $2\times2$ 
 line survey (see \citealt{esp13} and Marcelino et al., in prep).} We have found pointing errors between submaps typically of 2--4$''$, with a few cases of 
large errors of 7--10$''$ in RA under poor weather conditions. Besides the pointing errors in the 
observations, we found that pixels in the array are not perfectly aligned. We obtained a measurement 
of pixel misalignment in both HERA1 and HERA2 arrays, using $^{12}$CO and $^{13}$CO maps of IRC+10216 
(Cernicharo et al., in preparation). The errors found are between $0.1-2''$.
Further corrections on the intensity calibration were needed in order to account for the different 
response of each pixel within the HERA array. The different calibration between pixels is due to the 
fact that the image band rejection cannot be measured for each individual receiver in the array. 
Since this parameter will depend on the tuning, maps performed in different days will show different 
intensity calibration. Therefore, the data were scaled using a factor which depends on the particular 
submap and pixel in the array. This factor was obtained comparing the emission in each pixel with 
respect to the central one in the array, which was found to be the most uniform over the whole map. 
{  After all these corrections were applied, the final map was constructed using a pixel size of
5.5$''$ i.e. half the beam size}. The total number of positions in the map are 99415 after reduction, each with an on-source integration 
time of 10 seconds ($\sim$1 sec. at the most external parts of the map) corresponding with rms of 
$\sim$0.2 K (0.4--0.5 K at the map borders).

\begin{figure*}
\begin{center}
\includegraphics[width=18cm]{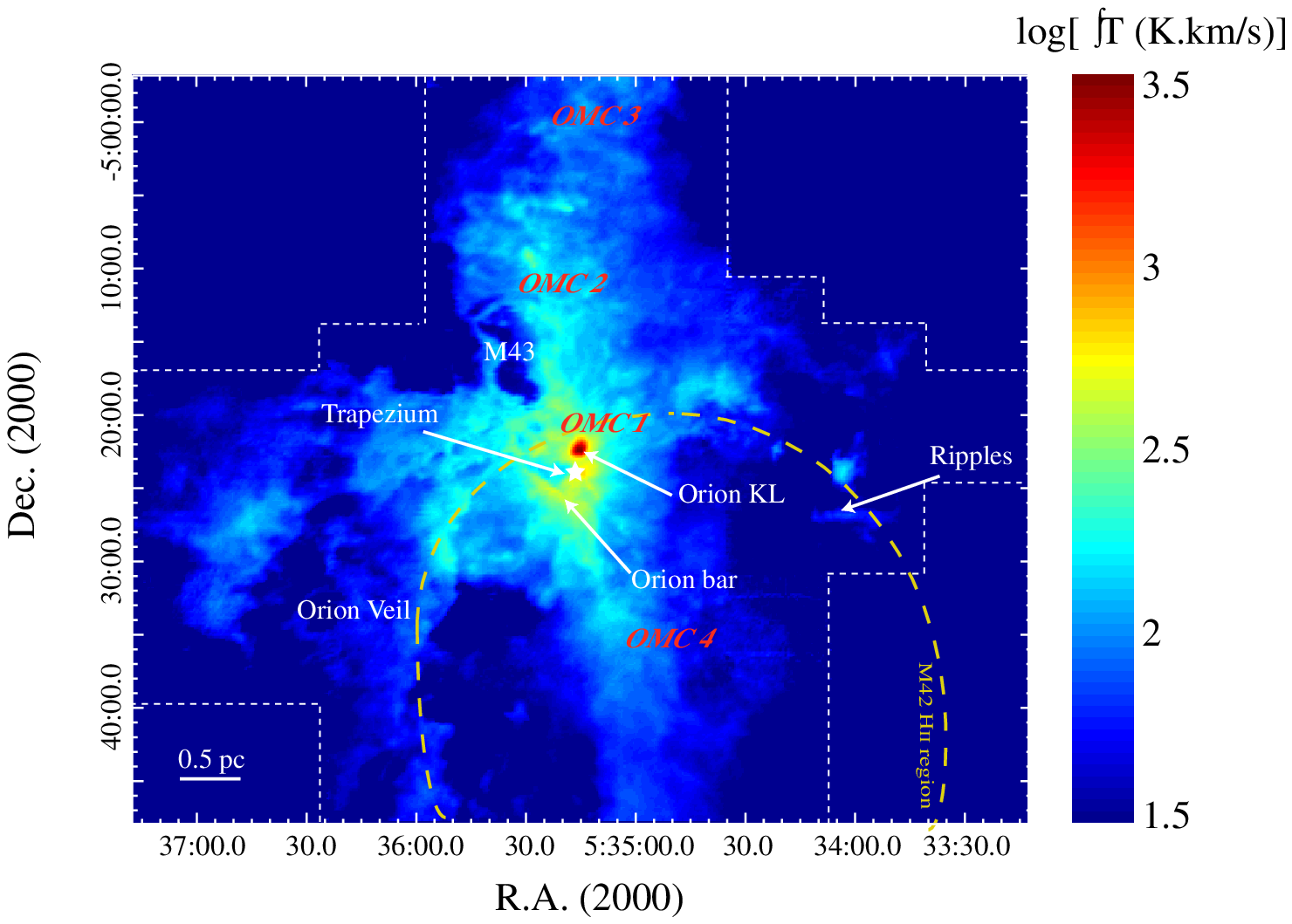}
\caption{{ IRAM 30m maps of the decimal logarithm of velocity integrated $^{12}$CO (2-1)  emission of the Orion molecular cloud. Distinctive objects within the Nebula are indicated. The yellow dashed line indicates the approximate limits of the HII region M42, based on the thermal Bremsstrahlung (free-free) emission
observed at 330 MHz by \citep{sub01}, while the white dashed line indicates the limits of the field of view.
See \url{http://userpages.irap.omp.eu/~oberne/Olivier_Berne/Data} for larger resolution versions of this image and data in FITS format.}
\label{map_tot1}}
\end{center}
\end{figure*}

\begin{figure*}
\begin{center}
\includegraphics[width=18cm]{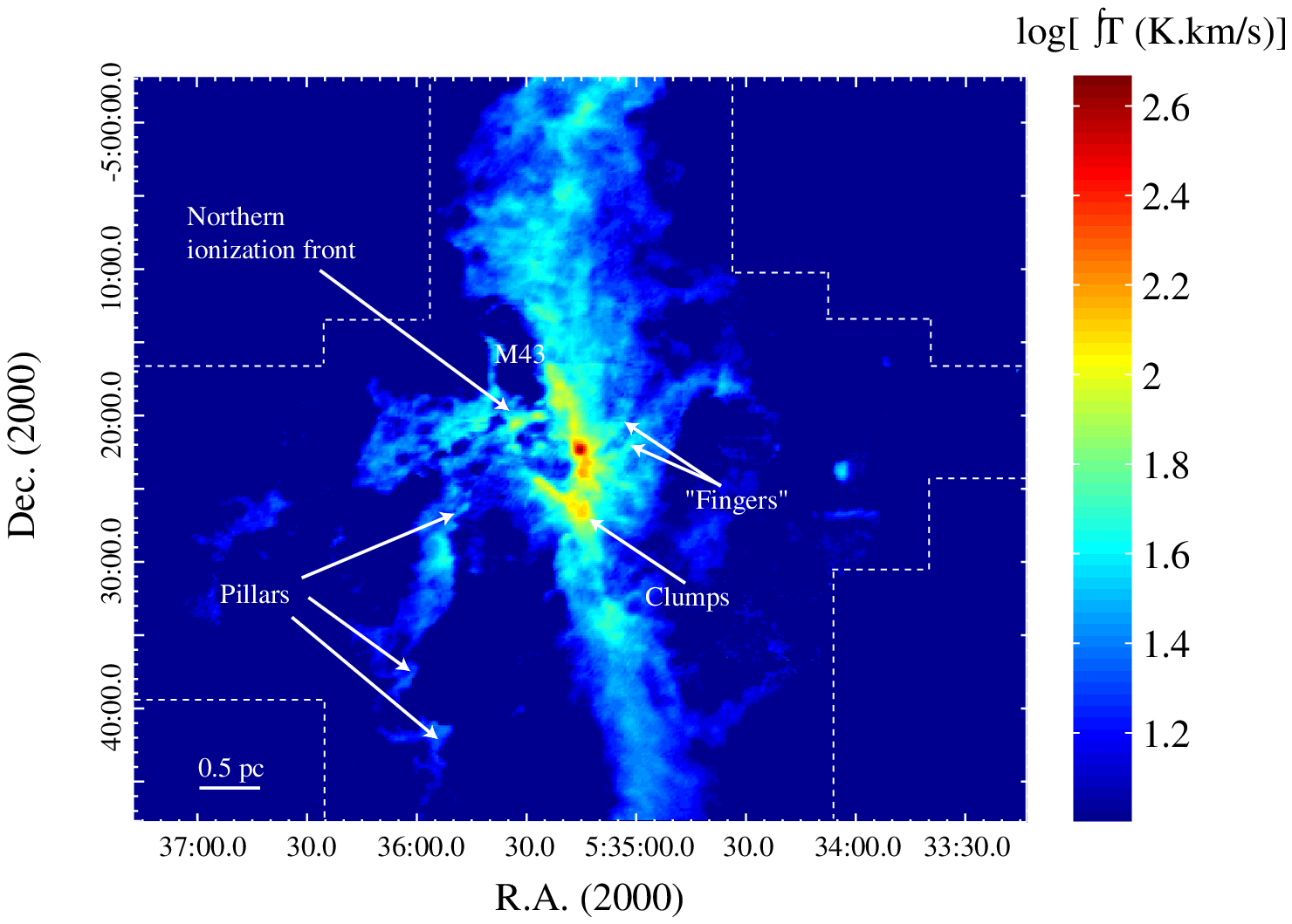}
\caption{{ IRAM 30m maps of the decimal logarithm of velocity integrated $^{13}$CO (2-1) emission of the Orion molecular cloud. Distinctive objects within the Nebula are indicated.
See \url{http://userpages.irap.omp.eu/~oberne/Olivier_Berne/Data} for larger resolution versions of this image and data in FITS format.}
\label{map_tot2}}
\end{center}
\end{figure*}

\section{Results}



Figures~\ref{map_tot1} and \ref{map_tot2} show the $^{12}$CO and $^{13}$CO $J=2-1$ integrated emission maps 
between $-$5 to 25 km~s$^{-1}$ . While $^{12}$CO shows a widespread emission across the observed region, 
the less abundant and optically thin isotopomer $^{13}$CO, delineates the densest regions, 
showing a very clumpy and filamentary structure. The bulk of the gas is mainly distributed 
along the ISF \citep{bal87}, clearly delineated by the densest gas seen in $^{13}$CO emission (Fig.~\ref{map_tot2}). 
This filament contains the Orion Molecular Clouds 1--4 (see below). The gas distribution extends East and West of the main 
filament, but not equally. Whilst emission spreads continuously to the East and southeast, showing 
numerous clumps and filaments, it is truncated sharply Southwest to the filament, coincident with 
a region where X-ray emission is observed (G\"udel et al. 2008). { Further West of this empty 
region where no CO emission was found}, a spherical clump and an elongated cloud (E-W) are observed. These features, 
are only seen at blue shifted velocities (see below), 
suggesting that they are being pushed toward the observed by a bubble of hot and ionised gas from 
massive stars. Indeed the elongated cloud is composed of several clumps (or ``ripples") whose geometry and velocity 
suggest they are the result of a Kelvin-Helmholtz instability produced by the expansion of the 
nebula (Bern\'e et al. 2010).

The Orion Molecular Cloud 1 (OMC-1), located behind the Orion Nebula, is the brightest region 
seen at the center of the filament and contains well-known features such as the BN/KL infrared 
nebula, the Orion-South region (Ori-S) and the Bar, which positions are indicated in Fig.~\ref{map_tot1}.


\begin{figure*}
\begin{center}
\includegraphics[width=13cm]{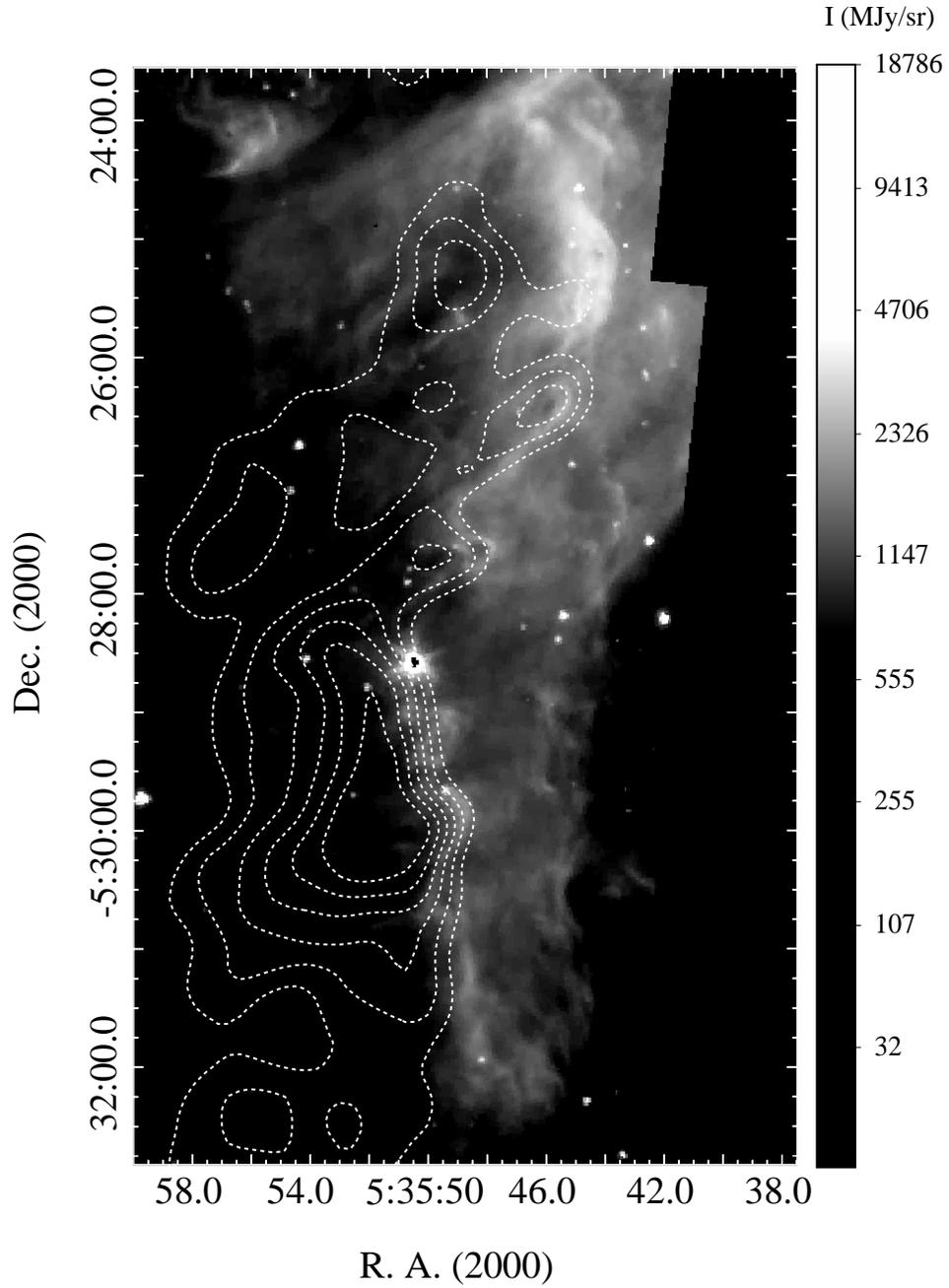}
\caption{ Map of the \emph{Spitzer}-IRAC 3.6 $\mu$m emission due to PAHs in the region of the pillars (grayscale). The dashed contours show the velocity integrated
$^{13}$CO (2-1) intensity. Contours start at 20 K.km.s$^{-1}$ and increase in steps of 4 K.km.s$^{-1}$. \label{fig_pillars}}
\end{center}
\end{figure*}

In contrast to the bright emission of the Bar, the KL nebula and Ori-S, there is a small region 
of weak emission which corresponds to the position of the Trapezium stars. This region 
is limited by the M42 HII region ionization fronts (the Northern Ionization Front and the Bar to 
the south), and by the molecular ridge to the West. It is easy to see, in the $^{12}$CO maps, the 
close relation of these bright features with the cone-like ionized structure of M42 observed in 
radio-continuum emission (Yusef-Zadeh 1990). The North Ionization Front (NIF) is formed by several 
filaments running sorthwest-southeast (see $^{13}$CO emission in Fig.~\ref{map_tot2}). The NIF is 
the counterpart of the obscured filament seen in the optical images of the Orion Nebula and known 
as the ``Dark Lane''. Further East, other clumps appear which follow the same direction as the 
NIF, indicating they might be related. In $^{12}$CO (Fig.~\ref{map_tot1}), 
low density gas East to the NIF is observed. This emission ends abruptly in a wall-like structure 
from which few thin filaments or streamers spread towards East, like tunnels blowing the molecular 
gas to the clumps seen at the eastern most part of the map. South to the NIF a filamentary structure 
going nearly North-South is seen (Fig.~\ref{map_tot2}), which should also 
be related to the expansion of the HII region (see \citealt{rod98}). Cometary shaped pillars are well 
visible in this region (Fig.~\ref{map_tot2}). The fact that { these} are pointing towards the Trapezium stars 
strongly suggest that these structures are the result of the selective photo-erosion of the Nebula, 
following the model proposed by \citet{rei83}. 
{This is illustrated in more details in Fig.~\ref{fig_pillars} where we overlay the 
$^{13}$CO emission and the \emph{Spitzer}-IRAC 3.6 $\mu$m image in this particular region.
While $^{13}$CO traces the inner molecular cloud, the IRAC image { is thought to trace the emission}
from UV heated polycyclic aromatic hydrocarbons (PAHs), present in the mostly atomic
gas at the surface of the molecular cloud. It is clear from Fig.~\ref{fig_pillars} that the PAH
emission enshrouds the molecular pillars and  peaks  systematically closer to the 
Trapezium stars as compared to CO. This stratification is classical in photon-dominated regions
(PDRs, see for instance \citet{tie93} for a classical example in the Orion Bar) and hence also favors the idea that selective photo-erosion is at the origin of these Pillars.}
Immediately to the West of Orion-KL, elongated structures or \emph{Fingers} \citep{rod92} are detected. These structures are believed to be the result
of the interaction of the molecular cloud with the extreme environment (especially the flow
of ionized gas, see discussion in \citealt{rod92}).

The northern part of the ISF shows strong $^{12}$CO emission  (Fig.~\ref{map_tot1}), which is separated in several clumps 
in the $^{13}$CO map (see Fig.~\ref{map_tot2}). This region correspond to the molecular 
clouds OMC-2 and OMC-3. Both regions are known to be very active in star formation since many 
young protostars and pre-main sequence stars have been detected in the infrared and submillimeter 
wavelengths, together with molecular outflows and jets (see \citealt{pet08}, for a review). 
Indeed, the turbulent $^{12}$CO emission in OMC-2/3 indicates the presence of such outflows (see 
section \ref{sec_kin}). The $^{13}$CO emission on the other hand, is very clumply (Fig.~\ref{map_tot2}), showing the densest cores 
where star formation is taking place. To the southeast of OMC-2, there is a cavity empty of molecular 
emission, clearly seen in the $^{13}$CO map (Fig.~\ref{map_tot2}) which corresponds 
to the HII region M43. {  This cavity, which in the optical is nearly spherical, is a small HII region illuminated
by the B0.5 star $\nu$ Ori \citep{thu78}}. M43 is limited in molecular emission by 
the ISF to the west, and a thin filament to the east that connects OMC-2 to the NIF. To the South, 
the ISF seems less clumpy and active in star formation than the northern filament (e.g. no outflow 
signatures observed). There is a local maximum located about 700'' south of Orion-KL, corresponding 
to the OMC-4 molecular cloud. This molecular core was identified in the CO maps of \citet{lor79} 
and in the SCUBA 450 and 850 $\mu$m emission observed by \citet{jon99}. In their 
submillimeter maps, this cloud is seen as a bright concentration of knots in a V-shaped structure. 


\section{Analysis}

\subsection{Column density map}

\subsubsection{Method}

\begin{figure*}
\begin{center}
\includegraphics[width=\hsize]{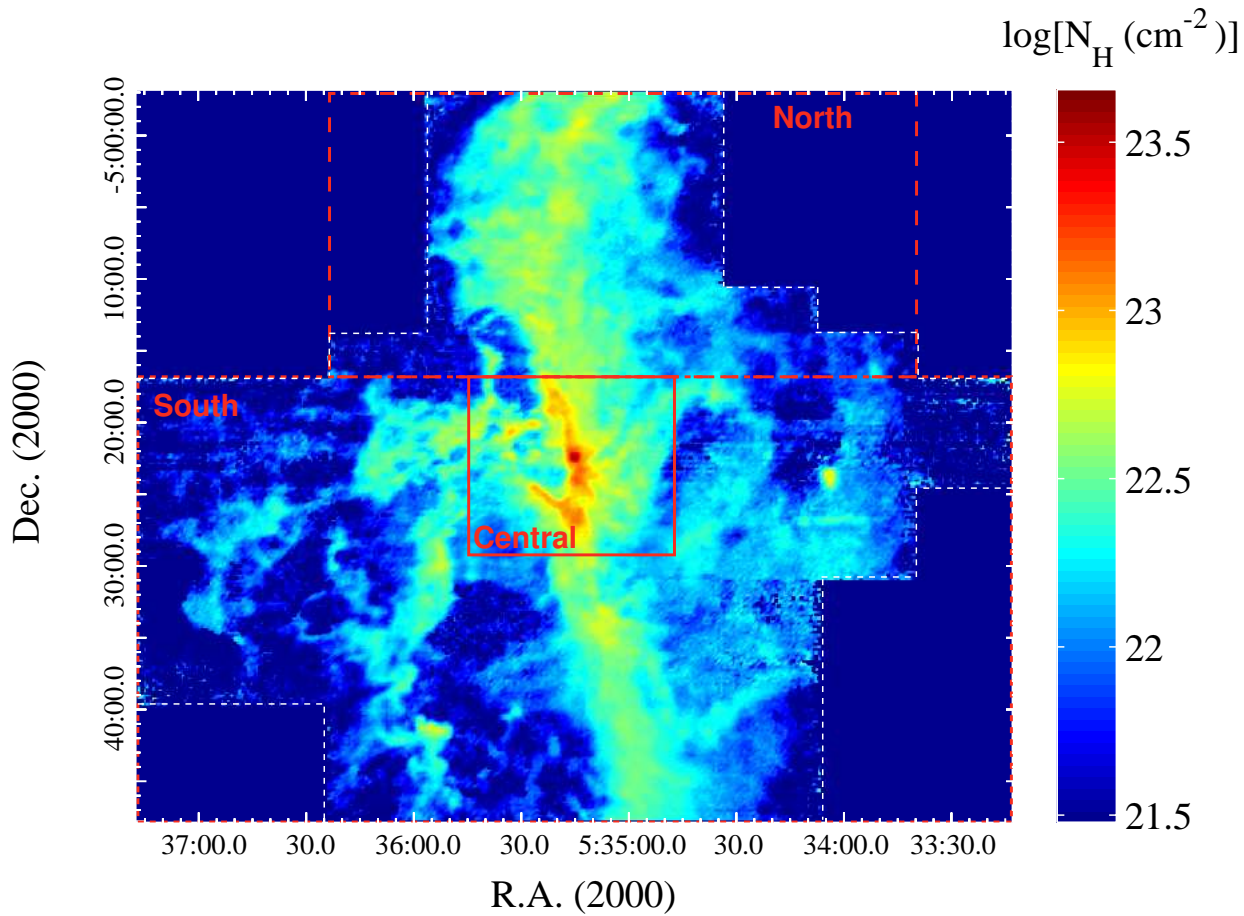}
\caption{Map of the decimal logarithm of column density $N_H$ in the Orion molecular cloud estimated from the $^{12}$CO(2-1) and $^{13}$CO(2-1) data (see text for details).
{ The threshold for this map is $N_H=3\times10^{21}$ cm$^{-2}$ {  corresponding to typical optical depths above which CO is expected to survive to 
photodissociation}. The borders of the sub-regions mentioned in the text and in Table~\ref{tab_con} are indicated with the red dashed boxes.
See \url{http://userpages.irap.omp.eu/~oberne/Olivier_Berne/Data} for larger resolution versions of this image and data in FITS format.}
\label{map_nh}}
\end{center}
\end{figure*}

{ Assuming  local thermodynamic equilibrium and optically thin emission (which is mostly the case for $^{13}$CO), 
the  $^{13}$CO column density can be obtained using an excitation temperature derived from the peak temperature of the $^{12}$CO line
(see, e.g. \citealt{gol99}). The  $^{13}$CO column density can be converted into a column density $N(H_{2})$ using the abundance ratio 
$[H_2]/[^{13}CO]=7\times10^5$.}
Of course, the derivation of column density in this manner is limited by the assumptions made, in particular
$^{13}$CO may become moderately optically thick in some regions of the Nebula with high column densities
(see e.g. \citealt{cas90}). In dense and cold regions, CO is also expected to be depleted on grains
and the $[H_2]/[^{13}CO]$ ratio may become larger than the one used. Both of these
effects will imply an underestimation of the total column density of the order of a few for dense regions.
In the rest of the nebula we expect our estimation to be relatively robust. 
We have computed $N_H$ on each pixel of the map (Fig. \ref{map_nh}).
Column densities derived in this way range between { a few times 10$^{18}$ and a few times
10$^{23}$ cm$^{-2}$}. We have mapped the points which correspond to column densities
below $3\times10^{21}$cm$^{-2}$, corresponding to the typical visual extinction
at which CO can survive (e.g. \citealt{dra11}). We found that these points correspond to {  noisy 
$^{13}$CO spectra i.e. with a peak brightness below 0.6 K, i.e. 3 times the typical sensitivity of 
$\sim 0.2$ K per channel}, where the column density cannot be 
evaluated {  with confidence}. Therefore we have thresholded our map at $3\times10^{21}$cm$^{-2}$ (Fig. \ref{map_nh}). 
{ We also define arbitrarly three subregions in the map presented in Fig.~\ref{map_nh}:
the North region, which contains OMC-2/3 and corresponds to the regions where
many low mass stars are present, the South region, which contains OMC4, the 
ripples and the cometary shaped pillars and corresponds to the parts of the cloud
which enshroud the M42 HII region, and finally,  the central region, corresponding to
the the most active and complex parts of the nebula, in which Orion-KL, the Bar, the molecular Fingers and the NIF
are found. }


\begin{figure*}
\begin{center}
\includegraphics[width=18cm]{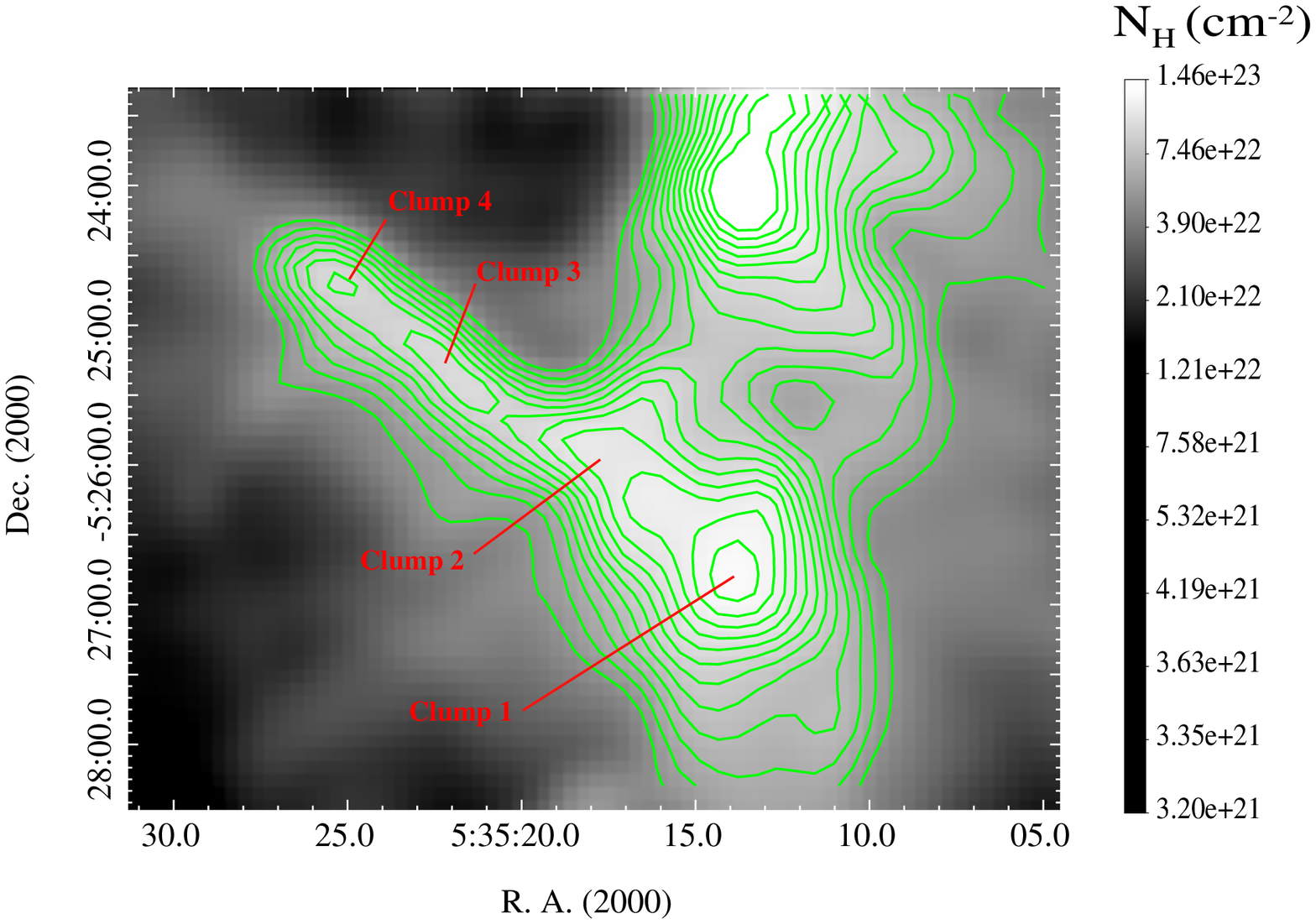}
\caption{Column density map of the Orion bar region (greyscale). The contours start at $5\times10^{22}$ cm$^{-2}$ and increase by steps of 
$2/3\times10^{22}$ cm$^{-2}$ up to $1.3\times10^{23}$ cm$^{-2}$. 
The four clumps identified in the Bar are indicated. 
\label{fig_orion_bar_Nh}}
\end{center}
\end{figure*}


\subsubsection{General properties}

The map of column density derived from the $^{13}$CO data is shown in Fig.\ref{map_nh}.
Moderate column densities of {  a few time} 10$^{21-22}$cm$^{-2}$  are found along the ISF.
The column density for the Orion-KL is of the 
order of 5$\times10^{23}$ cm$^{-2}$ (Fig.~\ref{map_nh}) 
close to the values reported recently by \citet{plu12}. 
{  Fig.~\ref{fig_orion_bar_Nh} presents a close-up view of the column density map in the 
Orion Bar region. The Bar has a typical maximum column density of $10^{23}$ cm$^{-2}$.
{  
This map suggests the presence of clumpy structures, and we have identified 4 specific clumps.
They correspond well in position (Table~\ref{tab_clumps}) and shape to the clumps seen in the $^{13}$CO (3-2) and C$^{18}$O (3-2)
maps of \citet{buc12} (see their Fig. 7). The good match between the positions of these clumps
in  $^{13}$CO (2-1), $^{13}$CO (3-2) and C$^{18}$O (3-2) indicates that the emission in these
transition is only moderately affected by optical depth effects in the Bar. 
We note that these clumps appear relatively large (diameter $>$ 14'' , see Table~\ref{tab_clumps}) and therefore 
differ from the arcsecond scale clumps proposed to exist by \citet{vdw96} 
and \citet{goi11}.In the case these mini-clumps also exist in the Bar, this would suggest that clumpy structures with 
different spatial scales coexist.}
The maximum column densities for these clumps are
9, 10, 11, and 13 $\times10^{22}$ cm$^{-2}$ for clumps number 4, 3, 2, and 1 respectively.
These values are comparable with the column densities derived for the Orion bar 
based on far infrared dust emission by \citet{ara12}.
Apart from the ISF and classical structures described above, other notables structures are
found with significant column densities: the ripples ({  a few times} 10$^{21}$cm$^{-2}$ ),
the clump just North of the ripples ({  a few times} 10$^{22}$cm$^{-2}$) as well as the wall of pillars at the 
edge the the Veil in the South-East. The molecular fingers also appear with high column densities 
(10$^{22-23}$cm$^{-2}$). The readers interested in the details of the structure of these
regions can find the fits file image of the column density map at \url{http://userpages.irap.omp.eu/~oberne/Olivier_Berne/Data.html}}

\subsection{Mass}
We can derive the {  total} mass of the cloud using:
\begin{equation}
M = \mu m_H \int_S{N(H_2) dS} \label{eq_mass}
\end{equation}
where $\mu = 2.8$ is the mean molecular weight and $m_H$ is the atomic hydrogen mass.
This integral is calculated on each pixel of the column density map, using the map 
pixel size (5''$\times$5'' at 414 pc) as the unit of surface. The total mass is then derived
 by summing all the integrals. Using this approach, we find a total mass of $\sim$7700 M$_{\sun}$,
 close to the value derived by \citet{cas90} using $^{13}$CO (1-0) for a similar field of view. 
 {  We computed the the mass of the integral shaped filament,
 using Eq.~\ref{eq_mass} but considering the field of view of \citet{buc12} (see Fig.~\ref{fig_region}). 
 This field is centered on Orion KL and follows the brightest emission of the
 ISF. With this field of view, we find a mass of 4300 M$_{\sun}$, in excellent
 agreement with the mass of 4290 M$_{\sun}$ derived by \citet{buc12} using $^{13}$CO (3-2)
 but smaller than the value reported by \citet{bal87} of 5000 M$_{\sun}$.
 Considering that these latter authors used a distance of 500 pc (instead of 414 pc here) 
 and a mean molecular weight of 2.6 (while we use a value of 2.8) to estimate the mass of the ISF,
 their value could be overestimated by a factor $(2.6\times500^2)/(2.8\times414^2)=1.35$ 
 as compared to ours. On the other hand, in their computation, \citet{bal87} consider a North-to-South 
 extent of about 1.5$^{\circ}$ for the ISF, compared to 0.85 $^{\circ}$ in our case. 
 Overall, we conclude that the mass we derive for the ISF is in very good agreement with 
 the value of  \citet{buc12}, where a direct comparison is possible. As compared the value of 
 \citet{bal87} the agreement is reasonable considering the uncertainty in the area used 
 to derive the mass. 
Since the total mass we derive for the cloud is about 7700 M$_{\sun}$, dense molecular gas present
in the ISF actually represents only about half of the cloud mass. Hence, a large fraction of the cloud
mass reside in more diffuse extended filamentary structures. Using the column density map, we 
have estimated that most of this mass in the \emph{diffuse} molecular gas  lies in regions of column 
densities lower than $3\times10^{22}$ cm$^{-2}$, or at a visual extinction $A_V$ below 10. The diffuse 
and filamentary molecular gas in these regions can be observed thanks to the high sensitivity obtained 
with the 30m telescope.} 
These results stress the importance of sensitive surveys in order to calculate accurate 
masses of molecular gas in nearby and extragalactic environments. 
{  Using the same method as above, we derive the masses of the clumps identified
in the Orion Bar (see Fig~\ref{fig_orion_bar_Nh}). We have assumed 
a circular geometry with radii given in Table~\ref{tab_clumps}.  The masses are found 
to range between 4.3 and 52 Solar masses ( Table~\ref{tab_clumps}). 
Condensations with similar masses are often observed around HII regions (see e.g. \citealt{deh09}),
and it is believed that they result from the compression of the molecular cloud due
to the expansion of the ionized bubble.}

\begin{table}
\caption{Properties of the clumps in the Orion bar}
\label{tab_clumps}
\begin{center}
\begin{tabular}{lccccc}
\hline \hline
Name & Position & Radius  & Mass  & Velocity dispersion & Virial paramater \\
Name & (R. A., Dec.) (deg.) & $R_{clump}$ ('') & $M_{clump}$ (M$_{\sun}$) &  $\delta V$ (km/s) &  $\alpha$\\
\hline
Clump 1 & (83.8075, -5.44827) & 46.3 & 52.1 & 3.03 &18.9\\
Clump 2 & (83.8192, -5.43942) & 23.4 & 13.6 & 2.08 &17.3\\
Clump 3 & (83.8431, -5.42102) & 23.8 & 11.3 & 2.11 & 21.8\\
Clump 4 & (83.8559, -5.41123) & 14.7 & 4.30 & 2.12 & 35.6 \\
\hline
\hline
\end{tabular}
\end{center}
\end{table}

\subsection{Kinematics}\label{sec_kin}

\subsubsection{Subtraction of the North-South velocity gradient}

{ 
As mentioned in \citet{bal87} (see their Fig. 5), the Orion A cloud is subject to a velocity gradient
over a scale of about 25 pc from North to South.This is also visible at the smaller scale which we study here and 
was reported as well in \citet{buc12} (see their Fig. 3 for a position-velocity diagram illustrating this effect).
When trying to study the small scale fluctuations of the velocity field, it is therefore useful 
to subtract this velocity gradient. Hence, both our $^{12}$CO and $^{13}$CO spectral 
cubes are corrected from a 0.7 km~s$^{-1}$~pc$^{-1}$ velocity gradient. 
This gives spectral cubes which are in the rest frame of the ISF, and can be used to study
spatially the fluctuations of the velocity field without contamination from the 
large scale \emph{falling backwards} effect. For these two cubes, we redefine 
$v=0$ at the peak of the line obtained by averaging the cube over the spatial dimensions.
}

\subsubsection{Kinematics from $^{12}$CO}

\begin{figure*}
\begin{center}
\includegraphics[width=17cm]{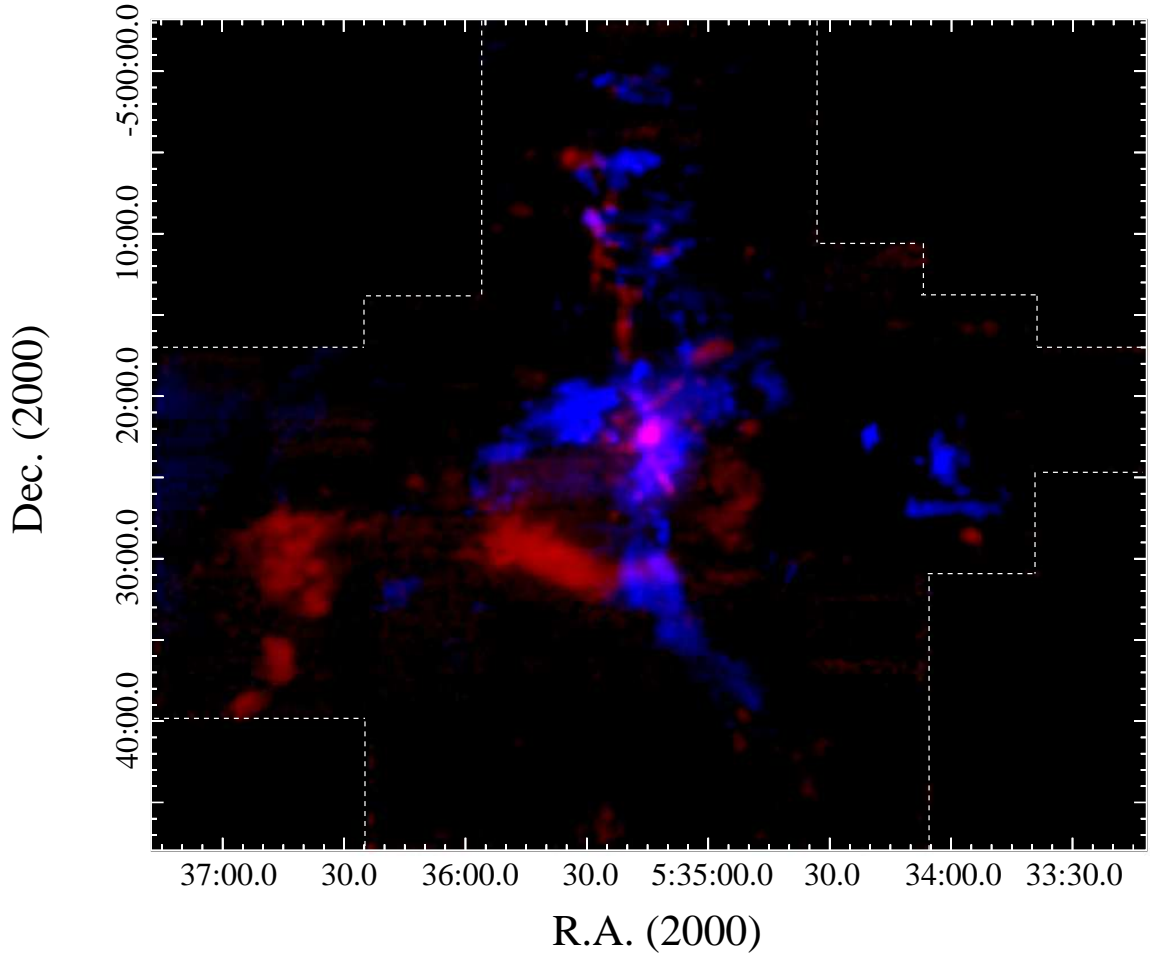}
\caption{{  Velocity integrated emission in the wings of the $^{12}$CO emission line, after subtraction of
the North-South velocity gradient to the spectral cube. Blue corresponds to 
velocities in the range $[-\infty, -4.8]$ km~s$^{-1}$ and red to velocities in the range $[+3.2, \infty]$ km~s$^{-1}$,
where the zero velocity correspond to the core velocity of the cloud  (see text for details).
\label{map_rgb_12}}}
\end{center}
\end{figure*}

\begin{figure}
\begin{center}
\includegraphics[width=16cm]{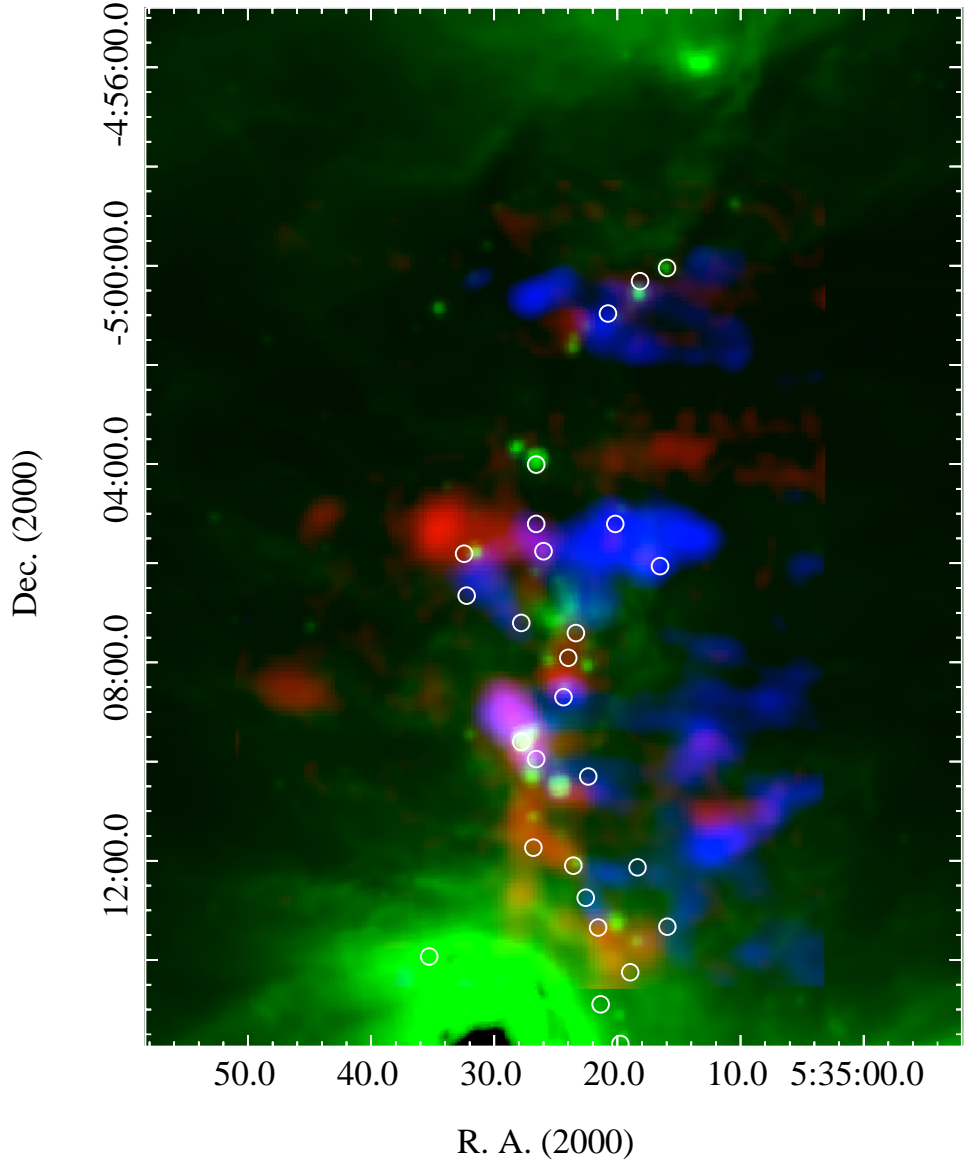}
\caption{{  Northern region of the nebula, showing the red and blue components of the $^{12}$CO emission (same as those of Fig.\ref{map_rgb_12}) as well as the
\emph{Spitzer}-MIPS 24 $\mu$m map in green. The far infrared clumps identified in the SCUBA 850 $\mu$m map of \citet{jon99} are indicated by the white circles.  
\label{outflows}}}
\end{center}
\end{figure}

{  
$^{12}$CO has the advantage to be bright but on the other hand this line quickly becomes 
optically thick. Hence, it is useful to study detailed spatial structures corresponding to emission
in the optically thin high velocity wings of this line. This is particularly suited for the study 
of outflows and jets emanating from protostars. Using the velocity gradient subtracted cube 
of $^{12}$CO, we compute two maps corresponding to the emission in the high velocity 
wings of this line. The ``blue" map (corresponding to blue shifted velocities in the frame of the ISF) 
is obtained by integrating the cube in velocity between $[-\infty, -4.8]$ km~s$^{-1}$.
The ``red" map instead is obtained by integrating over the $[+3.2, \infty]$ km~s$^{-1}$ range.
In Fig.~\ref{map_rgb_12}, we present an overlay of the ``blue" and ``red" maps.
In the South region, OMC-4 appears blue-shifted, mosty likely because it sits in front of the HII region, 
and is pushed towards the observer. To the east of OMC-4 is an arc shaped 
structure is clearly visible. This structure { had been} identified by \citet{lor79} and most likely represents the back of 
the cloud situated behind the HII region. The difference between the mean velocity of the blue and red shifted parts of the
cloud in the South region is of the order of $\Delta v\sim 10 $km~s$^{-1}$. With a radius of 1.8 pc for the HII region,
this implies a --short-- dynamical age of about $t_{dyn}\sim0.2$ Myrs for the expansion of the Nebula
due to the growth of the HII region.
The ripples also appear in blue, and hence seem to be part of a foreground cloud as was already proposed by \citet{bmc10}.
In the North region, $^{12}$CO looks more mixed/turbulent due to red and blue emission most likely arising due to the intense 
star formation and hence outflow activity in this region (e.g. \citealt{chi97, aso00, wil03}).
In Fig.~\ref{outflows}, we overlay the red and blue maps of $^{12}$CO to the MIPS
24$\mu$m image obtained by \emph{Spitzer}, focusing on the Northern region of the 
nebula only. In addition, we indicate the positions of the
far infrared clumpy structures that can be identified (by eye) in the SCUBA 850 $\mu$m map
obtained by \citet{jon99}. It clearly appears that numerous elongated blue/red shifted structures 
find their origin at positions of  24$\mu$m sources and/or FIR clumps, which are both
indicators of the positions of young stellar objects. We do not study these structures individually
since this has been done in great details by \citet{aso00} and \citet{wil03}.
Nevertheless, Fig.~\ref{outflows} suggests that most of the red and blue shifted structures 
seen in Fig.~\ref{outflows} correspond to protostellar outflows, and that the dynamical 
properties of the nebula in this region are likely to be affected, or perhaps dominated \citep{wil03}, by this activity.}

\subsubsection{Kinematics from $^{13}$CO}

\begin{figure*}
\begin{center}
\includegraphics[width=17cm]{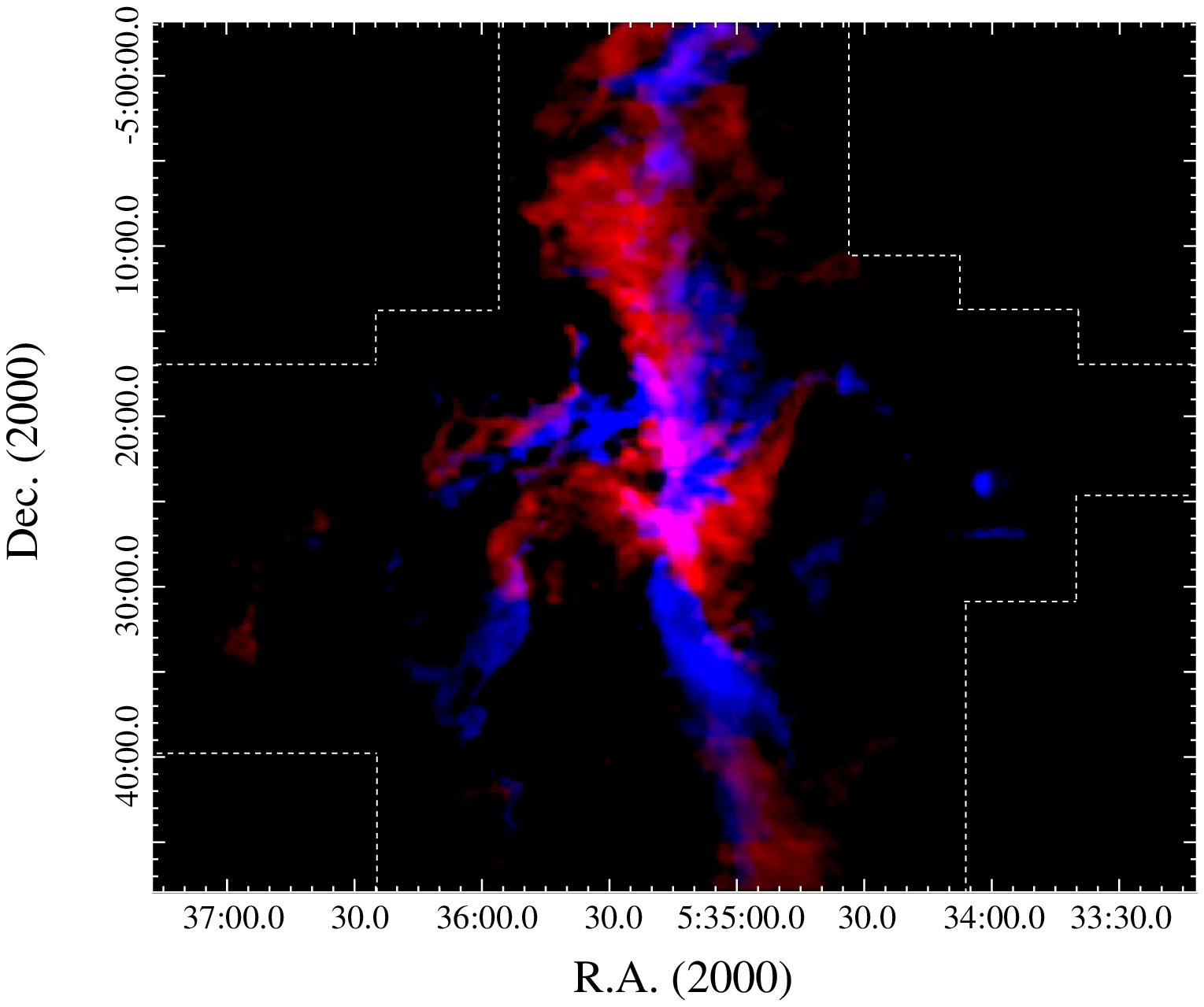}
\caption{{  Velocity integrated emission in the $^{13}$CO emission line, after subtraction of
the North-South velocity gradient to the spectral cube. Blue corresponds to 
velocities in the range $[-\infty, 0]$ km~s$^{-1}$ and red to velocities in the range $[0, +\infty]$ km~s$^{-1}$,
where the zero velocity correspond to the core velocity of the cloud  (see text for details).
\label{map_rgb_13}}}
\end{center}
\end{figure*}

{ 
As opposed to $^{12}$CO, $^{13}$CO has the advantage to be more optically thin but less bright. 
Hence, it is useful tracer to study the inner structure of the molecular gas. Using the velocity gradient subtracted cube 
of $^{13}$CO, we compute two maps corresponding to the emission in the blue and redshifted sides
of the line. i.e. the ``blue" map is obtained by integrating the cube in velocity between $[-\infty, 0]$ km~s$^{-1}$
and the ``red" map instead is obtained by integrating over the $[0, \infty]$ km~s$^{-1}$ range.
In Fig.~\ref{map_rgb_13}, we present an overlay of the corresponding ``blue" and ``red" maps.
Following the ISF from Orion KL, towards the North or the South, the filament successively 
appears in red and then blue, with a periodicity of the order of about 20'.
This pattern is compatible with a helical filament, which could 
be due to the structure of the large scale magnetic field which has been suggested to be 
helical (see \citealt{uch91, mat01, poi11, buc12}).
Whether this structure is at the origin of the formation of Orion KL (e.g. by the Parker instability as suggested by \citealt{shi91}) or resulting from the formation of Orion KL 
(e.g. acting as a drain of angular momentum as suggested by \citealt{uch91}) is unclear and
should be investigated in more details. In any case, this suggests that the magnetic field plays a dominant role 
in the evolution of the cloud. The comparative study between the kinematics of the cloud and the magnetic field combining 
the present CO data with the large -scale polarization data from the Planck satellite will be particularly useful to investigate
this subject.
}

\begin{figure}
\begin{center}
\includegraphics[width=16cm]{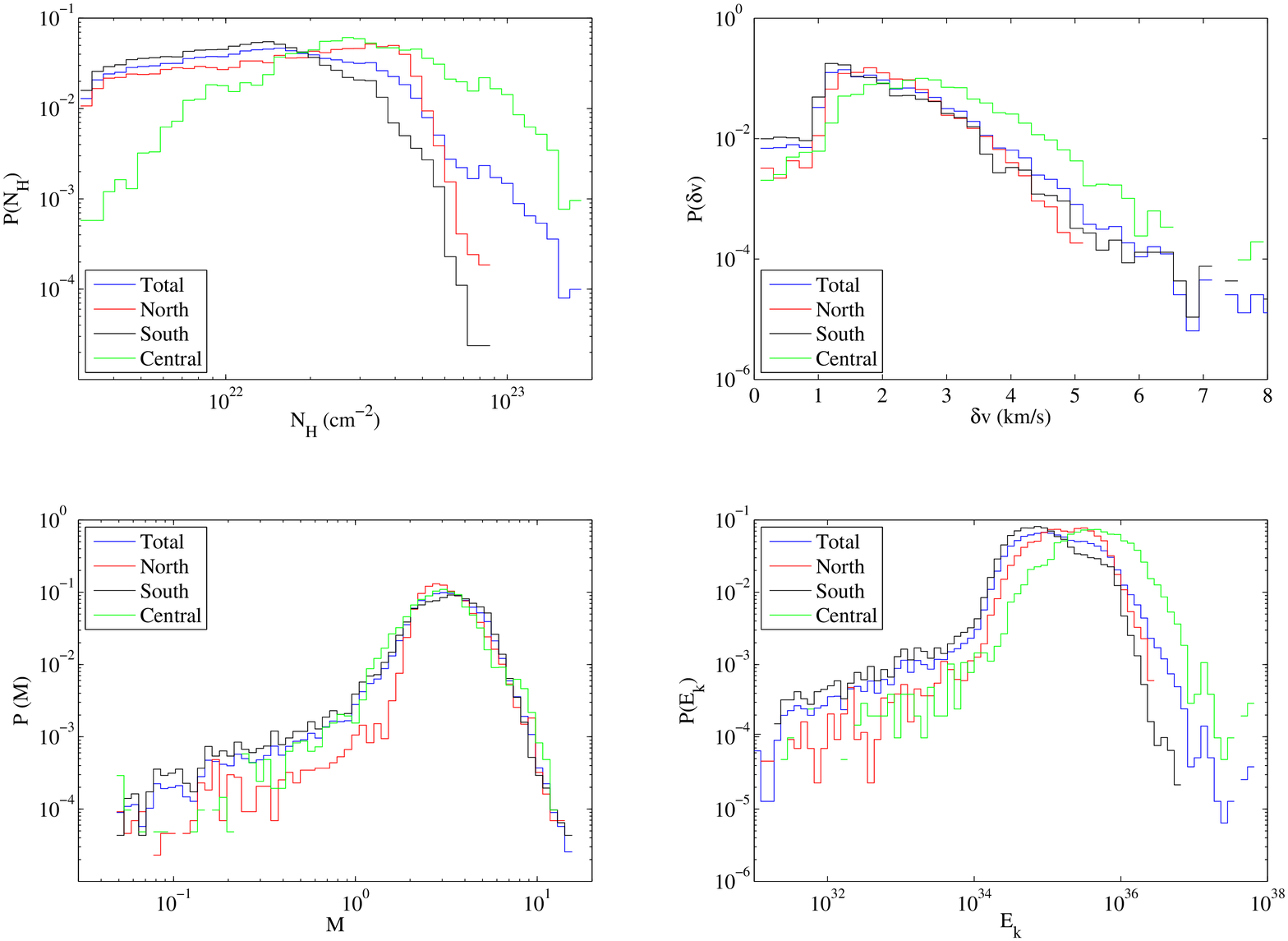}
\caption{Probability density functions of the per-pixel column density $N_H$, turbulent velocity dispersion $\delta v$, Mach number $\mathcal{M}$, and kinetic energy $E_k$
\label{histo}}
\end{center}
\end{figure}

\subsection{Energy balance}



Using the $^{13}$CO map we compute a map of the full width at half maximum (FWHM) of the lines and write this parameter $\delta v$. 
From this, we can derive a map of the turbulent kinetic energy in the gas defined by $E_{k}=1/2M \times \delta v^2$, with $M$ 
as defined in Eq.~\ref{eq_mass}.
The total kinetic energy of the cloud is $E_k=3.6\times10^{40}$ J. This value is slightly higher
than the kinetic energy derived by \citet{buc12} of $2.53\times10^{40}$ J, but, as for the mass,
this is probably because the field of view studied in the present paper is larger. 
In the central region,  {Orion KL contains $\sim 1.1\times10^{39}$ J of kinetic energy in the molecular gas.}
This number is one order of magnitude smaller than { that} derived by \citet{pen12}.
However, these authors consider the difference between the median velocity in the blue and red
wings of the CO (6-5) line to measure velocity dispersion. Therefore their number is not directly 
comparable to the one derived here. The Orion Bar contains $\sim 2.3\times10^{39}$ J, and the molecular 
fingers $\sim 3.7\times10^{39}$ J. The rest of the kinetic energy in the central region is distributed in various filamentary structures, 
for instance in the Northern ionization front filament. 
The North region holds about $8.9\times10^{39}$ J, a value comparable to the kinetic energy derived by 
\citet{wil03} using CO (1-0) observations of  $7.4\times10^{39}$ J, for the same region, and attributed to outflows. 

Assuming that most of the mass of the nebula lies in the ISF, and adopting a cylindrical geometry, we can calculate the gavitational energy
by $E_p=2GM^2/3R$ with R the radius of the cylinder. The CO images suggest a value
of $R=0.2^{\circ}=4.2\times10^{16}$m. This yields a gravitational energy $E_p=2.5\times10^{41}$
Joules, larger than the kinetic energy of the cloud. The thermal energy of the Orion molecular cloud,
$E_t=(3/2)k_BTM/\mu m_H=3.4\times10^{39}$ J, is an order of magnitude smaller than
the gravitational potential, adopting a temperature of 50 K for the molecular gas (the molecular gas would need to be 500 K
for the thermal energy to be comparable to the gravitational energy, which is unrealistic). The intensity of the magnetic field in the Orion Nebula has been 
measured by \citet{bro05} and \citet{abe04} and found to vary between 5 and 25 nT.
Adopting once more a cylindrical geometry, we can estimate magnetic energy by
$E_B=Rh^2B/8$, where $h$ is the North-to-South extent of the nebula (measured 
to be $h=1.9\times10^{17}$ m) and B the magnetic field strength. For the above mentioned values
of $B$, we find a range of magnetic energy $E_B=2-10\times10^{41}$ J. 

In summary, the magnetic energy is comparable, or larger, than the gravitational potential
and an order of magnitude larger than the kinetic energy. The thermal energy in the molecular 
cloud in negligible compared to these numbers. 

{  Now focusing on the Orion bar region, we consider the equilibrium for the four clumps
seen in the Orion bar (Fig. \ref{fig_orion_bar_Nh} ). For these four clumps, we derive the mean velocity dispersion $<\delta v>$,
from the map velocity dispersion $\delta v$ and inside a circular aperture with position and
radii given in Table 2. The values of $<\delta v>$ for each clump are given in Table 2. Using the values reported
in Table 2 for mass, radius and velocity dispersion, we can derive the virial parameter  defined as (e.g. \citealt{ber92})
$\alpha=5<\delta v>^2\times R_{clump}/GM_{clump}$ for each clump.

{ We find that all the $\alpha$ values
are large (Table 2), indicating that these clumps are subcritical and are confined by the external pressure or are evaporating.
Using the physical size (Table 2), column densities (Fig.~\ref{fig_orion_bar_Nh}) and a typical 
gas temperature of 30-50 K yields pressures of the order of a few $10^6$ K.cm$^{-3}$ for these clumps.
Assuming a magnetic field strength of the same order of magnitude as { that} derived by
\citet{abe04} (5-25 nT { i.e. 50-250 $\mu$G}) yields magnetic pressure also of the order of a few $10^6$ K.cm$^{-3}$
for the ambient medium. Since the magnetic field is expected to increase inside the clump where the density is 
higher, it  will be acting to support the clump against collapse.  Thus the cloud must be evaporating, unless the
ambient thermal pressure is greater than a few $10^6$ K.cm$^{-3}$. The results of \citet{abe04} indicate that the 
latter is not the case and hence these clumps are probably evaporating. Assuming evaporation at the sound speed yields
mass loss rates lower than a few $10^{-7}$ M$_{\sun}$/year, and these clumps will therefore live for timescales 
larger than that of the nebula.}

Overall, this suggest that the magnetic field plays a major  role in supporting the integral shaped
 filament \citep{bal87, fie00, buc12} as well a small scale structures such as  clumps.} 


\subsection{Probability density functions}\label{pdfs}
{
We derive the probability density functions (PDF) of the column density
$N_H$, turbulent velocity dispersion $\delta v$, and kinetic energy $E_k$. In addition, we derive the PDF of the 
turbulent Mach number $\mathcal{M}$, given by the ratio $\delta v / c_s$, where $c_s$ is the sound speed. 
$c_s$ is derived assuming the that the gas temperature is equal to the excitation temperature derived from the $^{12}$CO
line. For each one of these parameters, we compute the PDF for the whole region, but also for the 3 sub-regions 
identified in Fig.~\ref{map_nh}. The resulting PDFs are presented in Fig.~\ref{histo}. 

The column density PDF is found to be non-Gaussian and relatively flat. Column density histograms in star-forming regions
are known to exhibit power-law tails (see for instance recent examples in \citealt{kai11}). In our case the 
combination of a gaussian and a power law gives poor fits to the observed histogram. A collection of power laws 
can give a reasonably good fit, but this is at the expense of many arbitrary choices such as the number of power laws used.
It is interesting to note that similarly flat column density PDFs have been produced by MHD simulations of the turbulent saturation 
of the Kelvin-Helmholtz instability \citep{hen14}, which has been proposed as a source of small scale turbulence in the Orion molecular cloud
by \citet{ber12}. 
When comparing the PDFs of the three sub-regions, it clearly appears that the central regions hosts the largest column densities, 
present in the Orion Bar and Orion-KL. The North region also appears to have higher column densities than the more diffuse
South region. The North region shows a sharp decrease at a column density of $N_H\sim 4\times 10^{22}$ cm$^{-2}$,
perhaps corresponding to the limit of gravitational stability. Both the higher column density and presence of this sharp edge
are consistent with the fact that the North region is much more active in terms of star-formation than the South region. 

The PDFs of the velocity dispersion show a peak between 1 and 2 km/s, typical for molecular clouds and a log linear
tail towards larger velocities. The North and South regions show very similar $\delta v$ PDFs, while the PDF of the central
region is shifted towards larger velocities due to the intense kinetic activity of Orion-KL. 
The PDFs of the Mach number peak around 2-3, indicating that turbulence is highly supersonic in Orion. The PDF of the North region
appears sharper than in the central and South region, and peaks at a lower Mach number. This particularity is likely the result of the
low-mass star-formation activity i.e. linked to the fact that turbulence in this region is dominated by the outflow activity.

Finally, the last panel of Fig.~\ref{histo} shows the distribution of the kinetic energy. Logically,
the central region contains the pixels with highest kinetic energies, due to the presence or Orion KL. 
The North region also displays higher kinetic energies than the South region. This is most likely because of the
intense outflow activity in the North region, which is an efficient source of kinetic energy.
}

\subsection{Distribution of the feedback processes}

The distribution of the kinetic energy as well as other sources of energy
are summarized in Table 2.  In order to estimate the contribution from
different feedback process we classify the regions and their corresponding
kinetic energies either in the ``outflow-driven" or ``HII region-driven" feedback categories. 
The kinetic energy in the Orion bar is most likely the result
of the expansion of the HII region and there is very little outflow activity there.
On the contrary, Orion-KL is most likely dominated by outflows,
jets, or explosive motions, which all result from the feedback of massive protostars so we 
classify it in the outflow-driven feedback category. 
Following \citet{rod92} we consider that the kinetic energy in the molecular fingers
results from the interaction of the molecular cloud with the HII region. 
The situation is less clear for the various filamentary structures seen in the 
central region. The contribution of both the HII region and outflow
may be present so we consider that half of the feedback originates from outflows and the other
half from the HII region. 
In the South region which is directly in contact with the expanding HII region, 
and where large scale blue and red-shifted structures are seen, most of the kinetic energy
results form the feedback of the HII region. However there is known outflow activity in OMC-4, 
with a kinetic energy of $1.35\times10^{39}$ J  \citep{rod99}.
In the North, as suggested by Fig.~\ref{outflows} and in Sect.~\ref{pdfs}, the dynamical energy most likely
originates for a large part from outflows of low mass protostars (see also \citealt{wil03}) so we classify
this region in the outflow category.
All in all we find that the kinetic energy resulting from outflow activity $E_{OF}$ represents $1.5\times10^{40}$ J
while the kinetic energy due to interaction with the HII region $E_{i}$ represents $2.1\times10^{40}$ J.
This implies that 40\% of the kinetic energy injected in the cloud is provided by outflows 
and 60\% by the interaction with the HII region.

\subsection{Fraction of radiated energy converted into kinetic energy}

$E_{i}$ can be compared to the radiative energy injected in terms of ionizing photons by 
$\theta^1 C$.  Assuming that most of its luminosity lies in ionizing photons and using 
$L_*=2.1\times10^{5}L_{\sun}$, we find $E_*=L_* \times t_{dyn}\sim5\times10^{44}$ 
Joules. Most of this energy is transferred to the HII region \citep{fer01} in the form of 
thermal, kinetic and magnetic energy. With $E_{i}=2.1\times10^{40}$, this implies that
less than $\sim 0.01\%$ of the energy injected in starlight has been transferred in the molecular cloud in 
the from of kinetic energy.  This value is naturally small, because most of the energy of the 
HII region is radiated away. {  An additional fraction of kinetic energy may be present 
in the HII region (photo-evaporation flows, champagne flows see e.g. \citealt{och14}) but this has not been 
evaluated observationally so far.
Assuming a total lifetime for $\theta^1 C$ of $\sim$ 3 Myrs and a constant injection efficiency
implies that, in total, a few times $10^{41}$ Joules of kinetic energy will be injected in the ISM
due to the evolving HII region. This is at least one order of magnitude smaller than what can be injected 
by a supernova explosion \citep{tho98}, but is distributed over a much longer timescales,
and may thus have a profound impact on the evolution of the galactic ecosystem and on
triggered star formation.}

\subsection{On triggered star formation in Orion}

{ 
Triggered star-formation theories stipulate that the formation of a massive star and the
following expansion of an HII region can compress and fragment the surrounding molecular cloud,
resulting in the formation of a subsequent generation of low-mass stars. This process has
 been convincingly identified to be at play in the molecular shells around a number of 
 bubble shaped HII regions (see e.g. \citealt{deh10} for a recent overview). 
 Here on the other hand, we find that the molecular cloud directly in contact
 with the HII region in the South (OMC4, the Veil) shows lower column densities than 
 the Northern region (OMC 2-3). Such conditions disfavor star-formation, and indeed the 
 South region shows very few signs of star-forming activity although \citet{shi11} have 
 identified a few sub regions where star-formation could occur. Clearly,
 the North region contains much more protostars than the South region, and this suggest that
 the triggered star-formation due to the HII region is not effective in Orion. Given the young
 dynamical age we have derived (0.2 Myr), it is possible that triggering will occur later. Alternatively,
 an explanation could be that triggered star formation at the edge of expanding HII regions is
 only favored around well confined HII bubbles. Here in Orion, instead, the HII region
 may not be well confined and a champagne flow could be releasing the pressure, especially
 through a cavity towards the Eridanis super-bubble in the South-East \citep{gue08}. 
 In this case, the Orion HII region would already be in blister phase, and therefore the pressure
 driven expansion of the HII region is expected to be much reduced, as well as the 
 compression of the surrounding cloud and the triggering of star-formation.
}

\begin{table}
\caption{Derived parameters for the Orion molecular cloud}
\label{tab_con}
\begin{center}
\begin{tabular}{lr}
\hline \hline
\multicolumn{2}{c}{General parameters}\\
\hline
Total mass $M$							& 7700 $M_{\sun}$			 \\
Expansional velocity $\Delta V$				& 10 km~s$^{-1}$			 \\
Dynamical age $\tau_{dyn}$					& 0.2 Myrs			 \\
\hline
\multicolumn{2}{c}{Energy (Joules)}\\
\hline
Turbulent kinetic energy $E_k$					& $3.6\times10^{40} $ \\
\hline
South region (OMC-4)						& $ 1.3\times10^{40}$ \\
North region (OMC 2-3)						& $ 8.9\times10^{39}$ \\
Central region (OMC 1)						& $ 1.4\times10^{40}$ \\
$\ldots$ Orion bar							& $ 2.3\times10^{39}$ \\
$\ldots$ Orion KL							& $ 1.1\times10^{39}$  \\
$\ldots$ Molecular fingers							& $ 3.7\times10^{39}$ \\
$\ldots$ Filamentary structures in the central region		& $ 6.5\times10^{39}$ \\
\hline
Gravitational energy $E_p$					& $2.5\times10^{41}$ \\
Thermal energy $E_t$						& $3.4\times10^{39}$ \\
Magnetic energy $E_B$						& $2-10\times10^{41}$\\
Starlight $E_*$								& $5\times10^{44}$\\
\hline

\hline
\multicolumn{2}{ }\\
\end{tabular}
\end{center}
\end{table}

\section{Summary and conclusions}

{  We presented new large scale maps of the Orion molecular cloud in the
$^{12}$CO and $^{13}$CO (2-1) lines. The nebula appears filamentary
and turbulent. Using this dataset we derive the following parameters:
a dynamical age for the nebula $t_{dyn}\sim0.2$ Myrs, a total mass
of the nebula $M=7700 M_{\sun}$. About half of this mass resides 
in the dense integral shaped filament a the center of the nebula
while the other half resides in more diffuse and filamentary structures
distributed around the filament. We find a total  kinetic energy  $E_k=3.6\times10^{40}$ J
for the cloud, about 40\% of which emanates from the outflows
and the rest from feedback of the HII region. However, the relative importance
of the feedback processes depends on the considered sub-regions within the 
cloud. The North region of the cloud, (OMC-2/3) seems dominated by
feedbacks from outflows. In the central region (OMC-1), the feedback is shared between the protostellar 
outflows and feedback from the HII region. The South region, instead,
seems dominated by the feedback due to the expansion of the HII region. 
Overall, it seems that the feedback from the HII region, which is often 
invoked as the most important source in the triggering of star-formation is not
exceeding by a large number the feedback due to outflows, and therefore both mechanisms
may be important. In addition, we find that the triggering of star-formation 
by the expansion of the HII region seems inefficient in Orion.
We find that only about
$\sim 0.01\%$ of the energy radiated by $\theta^1 C$ has been converted into
kinetic energy in the molecular cloud. Over the lifetime of this star,
a few 10$^{41}$ Joules may be injected in the ISM. This is much smaller than
what can be injected by a supernova but can still have a significant influence on the
shaping of the ISM and the regulation of star-formation, since this occurs on larger timescales. 
Finally, we suggest that magnetic fields probably have an important role in the 
energy balance, at small and large scales.  We propose that the integral shaped filament 
has a helical shape, perhaps following the magnetic field lines, and is rotating 
about its main axis.}


\acknowledgments { We acknowledge the two referees of this paper who helped
improving the manuscript. We thank Paul Goldsmith for his comments.
The National Radio Astronomy Observatory is a facility of the National Science 
Foundation operated under cooperative agreement by Associated Universities, Inc.}

\bibliographystyle{apj}
\bibliography{biblio.bib}


\end{document}